\documentclass[%
 reprint,
 amsmath,amssymb,
 aps,
longbibliography
]{revtex4-1}

\usepackage[dvipdfmx]{hyperref}
\hypersetup{colorlinks=true, linkcolor=blue, urlcolor=dblue, citecolor=blue}
\usepackage[dvipdfmx]{graphicx}
\usepackage{amsmath} \usepackage{amssymb}
\usepackage{accents} \usepackage{mathrsfs} \usepackage{setspace}
\usepackage{color}
\usepackage{tcolorbox}
\definecolor{dgreen}{RGB}{00, 120, 00} \definecolor{dblue}{RGB}{00, 00, 180}
\definecolor{lgreen}{RGB}{46, 139, 87} 
\usepackage{ulem}

\newcommand{\bs}[1]{\boldsymbol{#1}} 
\newcommand{\tr}[1]{\mathrm{Tr}\left[#1\right]}

\newcommand{\ut}[1]{\undertilde{#1}} 
\newcommand{\ve}[0]{\varepsilon}

\begin{document}


\title{Quasiparticle Spectrum in Mesoscopic Superconducting Junctions
with Weak Magnetization}


\author{
S.-I. Suzuki$^{1,2}$, 
A.~A.~Golubov$^{2}$, 
Y.    Asano$^{3}$, and 
Y.    Tanaka$^{1}$}
\affiliation{
$^{1}$Department of Applied Physics, Nagoya University, Nagoya 464-8603, Japan, \\
$^{2}$MESA$^+$ Institute for Nanotechnology, University of Twente,
7500 AE Enschede, The Netherlands, and \\
$^{3}$Department of Applied Physics, Hokkaido University, Sapporo
060-8628, Japan, 
}


\date{\today}

\begin{abstract}
We theoretically investigate the effects of the weak magnetization on the
local density of states of mesoscopic proximity structures, where two
superconducting terminals are attached to a side surface of the
diffusive ferromagnet wire with a phase difference. When there
is no phase difference, the local density of states is 
significantly modified by the magnetization in both spin-singlet
$s$-wave and spin-triplet $p$-wave cases. When the phase difference is $\pi$, the local density of stets is less modified by the magnetization
compared with the in-phase case because of the destructive
interference of Cooper pairs.
\end{abstract}


\maketitle

\begin{figure*}[bt]
	\centering
  \includegraphics[width=0.96\textwidth]{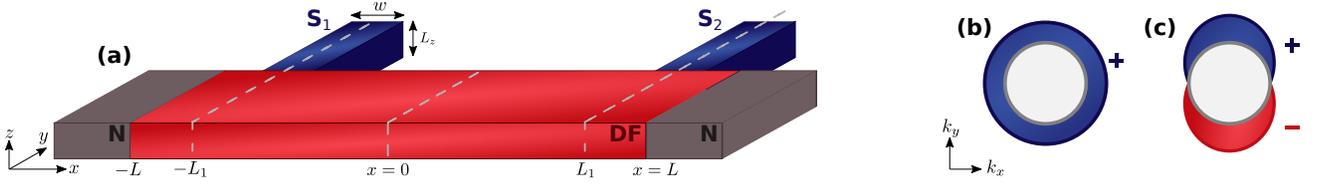}
	\caption{(a) Schematic of Volkov-Takayanagi 
	junction. The N/DF interfaces are located at $x=\pm L$. The
	barrier potential is present only at the S/DF interfaces. 
	The widths and the thickness of the
	wires are assumed $L_{y(z)} \ll \xi_0$. 
	The SCs with the phase difference $\delta \Phi$
	are attached to the DF at $x=\pm L_1$. 
	Schematics of the (b) $s$\,-wave and (c) $p$\,-wave pair potentials
	in momentum space. The magnetization is assumed weak $|\bs{M}| \sim
	\Delta$ and parallel to the $\bs{d}$-vector.}
	\label{fig:Sche}
\end{figure*}

\section{\label{sec:introduction} Introduction}


The proximity effect, the penetration of Cooper pairs, is a
characteristic phenomenon observed in superconducting junctions.
Cooper pairs penetrating into a non-superconducting material show
superconducting-like phenomena, for example, the suppression of the
local density of states (LDOS) at the Fermi level (zero energy) and
the screening of magnetic fields.  The penetration length of Cooper
pairs is limited to $\xi_T=(\mathscr{D}/2\pi T)^{1/2}$ with
$\mathscr{D}$ and $T$ being the diffusion constant and the
temperature.  Although $\xi_T$ is the characteristic length scale of
the proximity effect, Volkov and Takayanagi have shown that the
conductance of a two-superconductor (SC) junction [see
Fig.~\ref{fig:Sche}(a)] depends on the phase difference between the
two SCs even when the distance between two SC is much longer than
$\xi_T$ \cite{VT}. This effect is named the long-range phase-coherent
effect. 

The theory of the phase-coherence effect has been extended to
spin-triplet $p$-wave junctions \cite{Suzuki_PRB_2019}. The $p$-wave
SC is known to host Majorana bound states which is essential to
realize the quantum computation. The so-called Tanaka-Nazarov boundary
condition enables to study the energy spectrum of junctions with an
unconventional pairing \cite{Tanaka_Nazarov}. The Majorana zero-energy
peak (ZEP) in the LDOS has been found to vanish when the phase
difference is $\pi$ because of the destructive interference of Cooper
pairs. 

The ZEP can appear also in the junction of a diffusive
weak-ferromagnetic metal (DF) and an $s$-wave SC.  The energy shift
due to the magnetization $\bs{M}$ can result in an accidental ZEP when
$|\bs{M}|$ is comparable to the minigap size
\cite{Zareyan_PRL,Bergeret_PRB,Golubov_JETP,Yokoyama_PRB_2006}. In the
$p$-wave case, the magnetization known to change the Majorana ZEP to
the V-shaped LDOS\cite{Yokoyama_PRB_2007}. However, we have not known
how the accidental ZEP for the $s$-wave junction and the V-shaped LDOS
are modulated by the interference of Cooper pairs. The interplay
between the magnetization and the phase coherence has not been
elucidated yet. 

In this paper, we study the LDOS in a DF wire by solving numerically the quasiclassical Usadel equation. We consider the Volkov-Takayanagi (VT) junction as shown in Fig.~\ref{fig:Sche}(a), where two SCs are attached to the side surface of the DF with the phase difference $\delta \Phi$ . The order parameter is assumed the spin-singlet $s$-wave and triplet $p$-wave. The magnetization is assumed $\bs{M} = M \hat{\bs{z}}$ with $M \sim \Delta_0$.  In the $s$-wave case, the accidental ZEP caused by $M$ is flattened when $\delta \Phi = \pi$ due to the destructive interference among Cooper pairs injected from different SC wires. In the $p$-wave junction with $\delta \Phi = 0$, the LDOS is modified by $M$ from the ZEP to the V-shaped. When $\delta \Phi = \pi$, on the other hand, the LDOS is less modified by $M$ because the LDOS structure is less prominent due to the destructive interference. 


\section{System and Formulation}
\label{sec:KelUsa}

\subsection{Usadel equation}
In this paper, we consider the junctions of a DF 
where two superconducting (S) wires are attached to a side surface of
the DF as shown in Fig.~\ref{fig:Sche} [i.e., Volkov-Takayanagi (VT)
junctions]. 
Narrow S wires with the width $w$ are attached to the DF wire at 
$|x \mp L_1| < w/2$ and $y=0$ with an interface resistance $R_b$, where $w \ll \xi_0$  
with $\xi_0= \sqrt{\mathscr{D}/2\pi T_c}$. 
The DF is connected to 
clean normal-metal wires at $x = \pm L$ which are sufficiently narrow	and thin
in the $y$ and $z$ directions (i.e., $L_{y(z)} \ll \xi_0$). 

The Green's function in the DF obeys the Usadel equation\cite{Usadel}: 
\begin{align}
  & \mathscr{D} \bs{\nabla} \cdot
	\left( \check{g}^X \bs{\nabla} \check{g}^X \right)
	+ i \left[ \check{H}^X, \check{g}^X \right]_-
	= 0, 
	\\
	& \check{g}^X(\bs{r},\ve) = \left( \begin{array}{rr}
	     \hat{g} ^X &  \hat{f}^X \\[1mm]
	-\ut{\hat{f}}^X & -\hat{g}^X \\[1mm]
	\end{array} \right), 
	\label{eq:UsaUsa}
\end{align}
where $\mathscr{D}$ is the diffusion constant in the DF, $\check{g}^X(\bs{r},\ve)$
with $X=R$, and $A$ are the retarded and advanced
components of the Usadel Green's function, and $\check{H}^X$ is the
Hamiltonian-like matrix. 
In this paper, the accents $\check{\cdot}$ and $\hat{\cdot}$ means
matrices in particle-hole space and spin space.  The identity matrices
in particle-hole and spin space are respectively denoted by
$\check{\tau}_0$ and $\hat{\sigma}_0$.  The Pauli matrices are denoted
by $\check{\tau}_j$ and $\hat{\sigma}_j$ with $j \in [1,3]$.  The
Usadel equation is supplemented by the so-called normalization
condition: $\check{g}^X \check{g}^X = \check{1}$. 
Assuming the width of the
DF is much narrower than $\xi_0$, we can ignore the spatial variation
of the Green's function in the $y$ direction in the DF. Namely, one
need to consider a one-dimensional diffusive system where the Usadel
equation is reduced to 
\begin{align}
  & \mathscr{D} \partial_x 
	\left( \check{g}^X \partial_x \check{g}^X \right)
	+ i \left[ \check{H}^X, \check{g}^X \right]_-
	+ \check{S}^X \Theta_S(x)
	= 0, 
  \label{eq:Usadel-ori}
\end{align}
where the last term $\check{S}^X(x,\ve)$ represents effects of the S
wires and $\check{g}^X = \check{g}^X(x,\ve)$.  The source term $\check{S}^X(x,\ve)$ is reduced from the
boundary condition in the $y$ direction\cite{VT,Suzuki_PRB_2019}. 
The function $\Theta_S(x)$ is unity only beneath the S wires; 
$\Theta_S(x) = \Theta(w/2-|x-L_1|) + \Theta(w/2-|x+L_1|)$. 
The LDOS $\nu(x,\ve)$ and its deviation from the normal-state value 
$\delta \nu$ 
can be obtained from the $\check{g}^R$ and $\check{g}^A$ as 
\begin{align}
  & \nu(x,\ve) = \frac{\nu_0}{8}
	\tr{\check{\tau}_3 \left( \check{g}^R - \check{g}^A
	\right)}, 
	\\
  & \delta \nu(x,\ve) =
  \frac{ \nu (x,\ve)-\nu_0 } {\nu_0}. 
	\label{eq:dos}
\end{align}
where $\nu_0$ is the density of states per spin at the Fermi level 
in the normal states. 

In the presence of the weak magnetization, the matrix $\check{H}^X$ is
given by \cite{Bergeret_RMP_2005}: 
\begin{align}
  {\check{H}^X 
	= {\ve}^X \hat{\sigma}_0 \check{\tau}_3 
	  - M     \hat{\sigma}_3 \check{\tau}_0} , 
\end{align}
where $M$ is magnetization of the DF which is assumed much smaller
than the chemical potential \cite{Note}. 
The factor
$\ve^X$ depends on $X$: 
$\ve^R = \ve + i \gamma$ and 
$\ve^A = \ve - i \gamma$, 
where $\ve$ and
$\gamma$ being the energy and the depairing ratio due to inelastic
scatterings\cite{Dynes01} (i.e., Dynes formulation).
In this case, it is convenient to reduce the
$4 \times 4$ matrix into $2 \times 2$ one as 
\begin{align}
  & \mathscr{D} \partial_x
	\left( \tilde{g}^X_\alpha  \partial_x \tilde{g}^X_\alpha \right)
	+ i \left[ (\ve^X-\alpha M) \tilde{\tau}_3, \tilde{g}^X_\alpha \right]_-
	+ \tilde{S}^X \Theta_S(x)
	= 0, 
	\\
	& \tilde{g}^X_\alpha(x,\ve) = \left( \begin{array}{rr}
	   g^X_\alpha &  f^X_\alpha \\[1mm]
	  -f^X_\alpha & -g^X_\alpha \\[1mm]
	\end{array} \right), 
\end{align}
where the symbol $\tilde{\cdot}$ meaning a $2
\times 2$ matrix in spin-reduced particle-hole space, the factor
$\alpha = +1$ and $-1$ is for the majority and the minority spin,
respectively. Here we assumed $\delta \Phi = 0$ or $\pi$ by which
one can express $\ut{f}^X$ in terms of $f^X$. 
In what follows, we make $\alpha$ explicit only when
necessary. 
The Green's function can be simplified by the so-called 
$\theta$ parameterization\cite{Volkov_Physica_1993, Golubov_JLT_1988, Golubov_PRB_1997}:
  $\tilde{g}^X 
	 =
    \tilde{\tau}_3 \cosh \theta
  +i\tilde{\tau}_2 \sinh \theta$, 
where we omit the index $X$ from $\theta = \theta^X(x,\ve)$. The Usadel equation is reduced to 
\begin{align}
  & \mathscr{D} \frac{\partial^2 \theta_\alpha}{\partial {x}^2}
	+ 2 i (\ve^X - \alpha M) \sinh \theta_\alpha 
	+ \Theta_S(x) {S}(x,\ve) = 0. 
  \label{eq:Usadel-th} 
\end{align}
The Usadel equation \eqref{eq:Usadel-th} is supplemented by the
boundary conditions. 
The boundary conditions for $\delta \Phi = 0$ 
is given by 
$
             \theta(x,\ve) |_{x=\pm L} = 0$ and 
$ \partial_x \theta(x,\ve) |_{x=0    } = 0
$, 
whereas that for $\delta \Phi = 0$ is 
$  \theta(x,\ve) |_{x=\pm L} = 0$ and 
$  \theta(x,\ve) |_{x=0    } = 0$. 
%
\begin{figure*}[t!]
	\centering
  \includegraphics[width=0.98\textwidth]{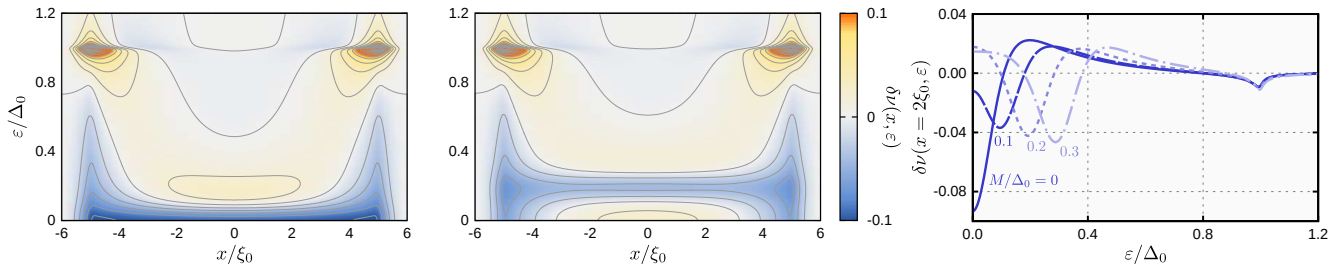}
	\caption{(Left and center) Deviation from the LDOS 
	$\delta \nu$ in the $s$-wave junction \textit{without} the phase difference. 
	The magnetization is set to $M=0$ in (Left) and
	$M=0.2\Delta_0$ in (Center). The lengths are set to $L=6\xi_0$ and
	$L_1=5\xi_0$. The ZEP appears in
	(Center) as a result of the energy shift by $M$. The sharp peaks at
	$\ve=\Delta_0$ and $|x|=L_1$ are the coherence peaks by the proximity
	effect. (Right) Effects of
	$M$ on $\delta \nu$ at $x=2\xi_0$. The parameters are set to $z_0 =
	1$, $\gamma_B = 1$, $L_1 = 5\xi_0$, $L = 6\xi_0$. }
	\label{fig:res_s0}
\end{figure*}
\begin{figure*}[t!]
	\centering
  \includegraphics[width=0.98\textwidth]{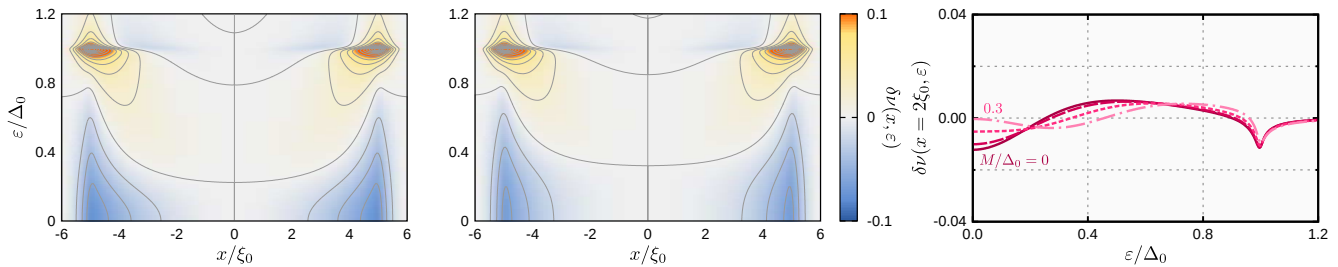}
	\caption{Deviations $\delta \nu$ in the $s$-wave junction with $\delta \Phi = \pi$. 
	The results are plotted in the same manners as in Fig.~\ref{fig:res_s0}.}
	\label{fig:res_s1}
\end{figure*}

\subsection{Effects of superconducting terminals}

The source term $S(x,\ve)$ in Eq.~\eqref{eq:Usadel-th} represents the effect of the S arms\cite{Volkov_Physica_1993,Golubov_PRB_1997}. The
typical boundary conditions\cite{Zaitsev_1984} are
no longer available for unconventional pairings. Therefore, one must employ the
so-called Tanaka-Nazarov
condition \cite{Tanaka_Nazarov}, an extension
of the circuit theory \cite{Nazarov_PRL_1994}. The boundary condition in the $y$ direction
is reduced to the source term ${S}$: 
\begin{align} 
  & \left. \frac{d \theta}{d \bar{y}} \right|_{y=0} =
  \frac{R_N}{R_b \bar{L}_y} \langle F \rangle_\phi, 
	\\
	& F = \frac{ -2 T_N (f_S \cosh \theta_0 - g_S \sinh \theta_0)}
	{(2-T_N)\Xi    +T_N (g_S \cosh \theta_0 - f_S \sinh \theta_0)},
  \label{eq:Tanaka-Nazarov} 
\end{align} 
where $R_N = \rho_N L_y / L_z w$, $\rho_N$ and $R_b$ are the specific
resistance of the DF and the interface resistance at the DF/S interface,
$\bar{y} = y / \xi_0$, and $\bar{L}_y = {L}_y / \xi_0$. 
The angle-dependent function
$T_N(\phi) =
\cos^2\phi/ (\cos^2\phi + z_0^2)$ is the transmission coefficient of
the N/N interface with the $\delta$-function
barrier potential $\hbar v_F z_0 \delta(y)$, $\phi$ is
the angle of the momentum measured from the $k_y$ axis, and
$\theta_0(x) = \theta(x)|_{y=0}$. The angular bracket means the angle average: 
$ \langle \cdots \rangle_\phi
	\equiv 
	\big( \int_{-\pi/2}^{\pi/2} \cdots \cos \phi~d\phi \big)
	\big( \int_{-\pi/2}^{\pi/2} T_N    \cos \phi~d\phi \big)^{-1}. 
$
The functions $g_S$ and $f_S$ can be obtained from the Green's
functions in a homogeneous ballistic SC:
\begin{align}
  & g_S = g_{S+} + g_{S-}, \\
  & f_S = 
	\left\{ \begin{array}{ll}
		f_{S+} + f_{S-}                  & \text{~for singlet SCs, } \\[1mm]
	  i(g_{S-} f_{S+} - g_{S+} f_{S-}) & \text{~for triplet SCs, } \\
	\end{array} \right. 
\end{align}
where	
$g_{S\pm}(\phi) = {\ve}       /{\sqrt{ \ve^2 - |\Delta_\pm|^2}}$,
$f_{S\pm}(\phi) = {\Delta_\pm}/{\sqrt{ \ve^2 - |\Delta_\pm|^2}}$, 
$ \Xi = 1 + g_{S+} g_{S-} - f_{S+} f_{S-} $, 
the symbol $X$ has been omitted, and 
$\Delta_+(\phi) = \Delta_-(\pi-\phi)$. The pair potential
depends on the pairing symmetry of the SC: 
$\Delta_+(\phi) = \Delta_0$ for an $s$\,-wave SC and 
$\Delta_0 \cos \phi$ for a $p$\,-wave one, 
where $\Delta_0 \in \mathbb{R}$ characterizes the amplitude of the
pair potential and the $\bs{d}$-vector is assumed $\bs{d} \parallel
\bs{M} \parallel \bs{z}$. {In this paper, the spin-orbit coupling is not
taken into account for simplicity\cite{Note}.}
The boundary condition \eqref{eq:Tanaka-Nazarov} is transformed into 
the source term: 
\begin{align}
  S(x,\ve) = 
	\frac{\mathscr{D}}{\xi_0^2} \gamma_B^{-1}
	\langle F(x,\ve,\phi) \rangle_\phi, 
\end{align} 
where $\gamma_B = R_b \bar{L}_y^2 / R_N$ is the dimensionless 
parameter. The parameters $\gamma_B$ and $z_0$ are 
independent. The function $R_b$ is a function of $z_0$ because it
determines $T_N$, whereas $\gamma_B$ determines how strong the
proximity effect is. 

\begin{figure*}[tb]
	\centering
  \includegraphics[width=0.98\textwidth]{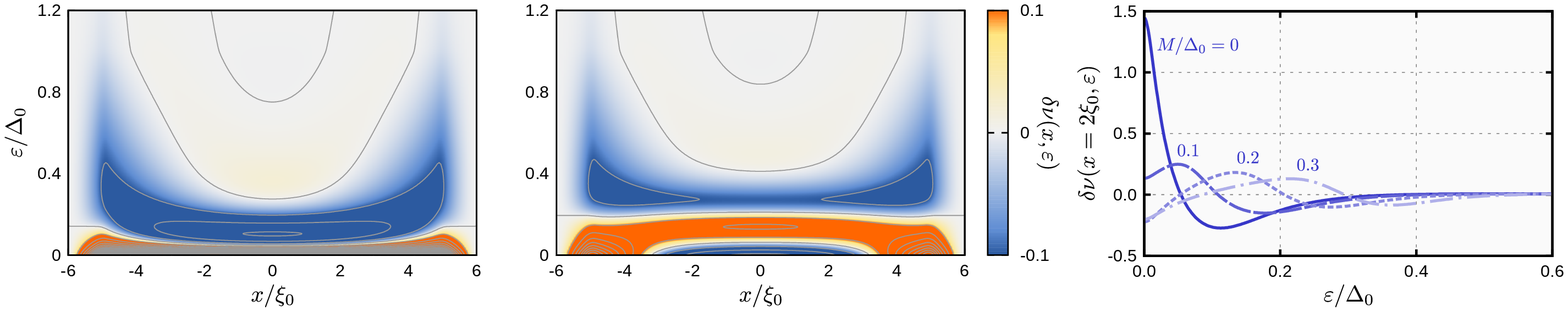}
	\caption{ Deviations $\delta \nu$ in the $p$-wave junction with
	$\delta \Phi = 0$.  The results are plotted in the same manners as
	in Fig.~\ref{fig:res_s0}.}
	\label{fig:res_p0}
\end{figure*}
\begin{figure*}[tb]
	\centering
  \includegraphics[width=0.98\textwidth]{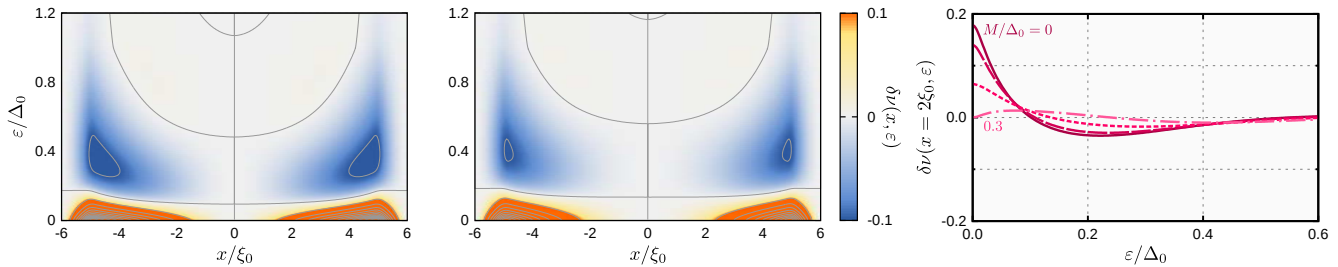}
	\caption{
	Deviations $\delta \nu$ in the $p$-wave junction with $\delta \Phi = \pi$. 
	The results are plotted in the same manners as in
	Fig.~\ref{fig:res_s0}.}
	\label{fig:res_p1}
\end{figure*}



\section{Numerical results}

We first show the results for the spin-singlet $s$-wave junctions 
in Figs.~\ref{fig:res_s0} and \ref{fig:res_s1} 
where the phase difference is set to $\delta \Phi = 0$ and $\pi$ 
respectively. 
The parameters are set to $z_0 = 1$, $\gamma_B = 1$, $L_1 = 5\xi_0$,
and $L = 6\xi_0$ throughout this paper. 
%
Beneath the S wire, the proximity effect results in the peaks at
$\ve=\Delta_0$ and $|x|=L_1$ (i.e., the coherence peak) and the
zero-energy dip. 
As shown in the left
panel of Fig.~\ref{fig:res_s0}, the zero-energy dip appears 
between the two SC terminals when $\delta \Phi = 0$ and $M=0$. On the other hand, the low-energy
spectrum is a peak \cite{Yokoyama_PRB_2006} when $M=0.2\Delta_0$ as in the center 
panel of Fig.~\ref{fig:res_s0}. The peak becomes higher as with
increasing the distance from the SC terminal. The energy shift by $M$, which depends
on the quasiparticle spin, results in 
this \textit{accidental} ZEP as shown in the right
panel of Fig.~\ref{fig:res_s0}, where we show $\delta \nu$ at $x=2\xi_0$. 

When $\delta \Phi = \pi$, the Cooper-pair interference is destructive. 
Therefore the pair amplitude from each SC terminal perfectly
compensate each other, 
leading $\delta \nu|_{x=0} = 0$ as shown in the left and center panels of 
Fig.~\ref{fig:res_s1}. Moreover, comparing Figs~\ref{fig:res_s0} and 
\ref{fig:res_s1}, we see that the zero-energy dip and ZEP become
less prominent compared with those with $\delta \Phi = 0$. Therefore,
the destructive interference is concluded to diminish the
zero-energy dip and the ZEP by $M$ in the DF. 

The results for the spin-triplet $p$-wave junctions are shown 
in Figs.~\ref{fig:res_p0} and \ref{fig:res_p1},
where the phase difference is set to $\delta \Phi = 0$ and $\pi$ 
respectively. 
The topologically-protected ZEP characterises the
spin-triplet $p$-wave junction as shown in the left panel of
Fig.~\ref{fig:res_p0} \cite{RG, SIS_Majo}. This ZEP is caused by the induced odd-frequency
Cooper pairs \cite{Tanaka_PRL_2007,Suzuki_PRB}. The ZEP is robust against the weak
magnetization \textit{beneath} the SC wire (i.e., $|x| \sim L_1$) 
but fragile \textit{between} them as shown in the center panel of 
Fig.~\ref{fig:res_p0}. The LDOS deviation $\delta \nu$ at $x=2\xi_0$ is
shown in the right panel of Fig.~\ref{fig:res_p0}. The LDOS changes
from the peak to the dip with increasing $M$ because of the
spin-dependent energy shift. The split LDOS peak appears V-shaped at
the low energy when
$M = 0.1\Delta_0$ and $0.2\Delta_0$. In the $M \gg \Delta_0$ limit,
$\delta \nu$ reaches to zero. 

When $\delta \Phi = \pi$, the ZEP around $x=0$ vanishes
due to the deconstructive interference as discussed in
\cite{Suzuki_PRB_2019}. As a result, the peak splitting
becomes much less prominent as shown in the
center panel of Fig.~\ref{fig:res_p1}, where the peak height is less
than 0.2. The LDOS deviation at
$x=2\xi_0$ is shown in the
right panel of Fig.~\ref{fig:res_p1}. Even when $M=0$, the LDOS peak
is much smaller compared with that for $\delta \Phi =0$. The
magnetization $M$ makes this LDOS peak much smaller.

\section{Summary}
We have investigated the effects of the weak magnetization on the LDOS
of mesoscopic proximity structures, where two SC terminals with a
finite phase difference $\delta \Phi$ are attached to the side
surface of the DF. 

In spin-singlet $s$-wave junctions with $\delta \Phi = 0$, the
accidental ZEP of the LDOS appears due to the energy shift by the
magnetization, whereas the LDOS structures become less prominent when 
the phase difference is $\delta \Phi = \pi$ because of the destructive interference of Cooper
pairs injected from different SC terminals. 

In spin-triplet $p$-wave junctions are characterised by the
topologically-protected ZEP in the LDOS, corresponding to the Majorana
bound state. When $\delta \Phi = 0$, the
LDOS is split and becomes V-shaped at the low energy by the
magnetization. On the contrary, when $\delta \Phi = \pi$, the effects
of the magnetization is not significant because the ZEP is much
smaller even when $M=0$ due to the destructive interference. 

\section*{Acknowledgments}
This work was supported by Grants-in-Aid from JSPS for Scientific
Research on Innovative Areas ``Topological Materials Science''
(KAKENHI Grant Numbers JP15H05851, JP15H05852, JP15H05853 and JP15K21717), 
Scientific Research (B) (KAKENHI Grant Number JP18H01176), 
Japan-RFBR Bilateral Joint Research Projects/Seminars number 19-52-50026,  
JSPS Core-to-Core Program (A. Advanced Research Networks).
A.~A.~G. acknowledges supports by the European Union
H2020-WIDESPREAD-05-2017-Twinning project ``SPINTECH'' under grant
agreement Nr.~810144. 

\clearpage


\begin{thebibliography}{99}
\bibitem{Note}
  {
	Although odd-parity superconductivity is often found in $5f$-electron
	systems such as uranium compounds where the spin-orbit coupling
	(SOC) plays roles, we ignore the effects of SOC for simplicity. 
	Therefore, in this paper, the magnetization is assumed to interact
	simply with electron spin.}
\end{thebibliography}
\end{document}